\documentclass[prl,twocolumn,showpacs,preprintnumbers,amsmath,amssymb]{revtex4}
\usepackage{graphicx}
\usepackage{dcolumn}
\usepackage{bm}
 
\begin{document}

\title{ Robust Quantum Computation with Quantum Dots }
\author{ C. Stephen Hellberg }
\email[Email: ]{ mylastname@dave.nrl.navy.mil}
\affiliation{ Center for Computational Materials Science,
Naval Research Laboratory, Washington, DC 20375}
\date{\today}
\begin{abstract}
Quantum computation in solid state quantum dots faces
two significant challenges:
Decoherence from interactions with the environment and the difficulty
of generating local magnetic fields for the single qubit rotations.
This paper presents a design of composite qubits to overcome
both challenges.
Each qubit is encoded in the degenerate ground-state 
of {\em four} (or {\em six}) electrons
in a system of {\em five} quantum dots arranged in a two-dimensional pattern.
This decoherence-free subspace
is immune to both collective and local decoherence,
and resists other forms of decoherence, which must raise the energy.
The gate operations for universal computation are simple
and physically intuitive, and are controlled by modifying
the tunneling barriers between the 
dots---Control of local magnetic fields is not required.
A controlled-phase gate can be implemented in a single pulse.
\end{abstract}

\maketitle

A quantum computer 
with a sufficient number of quantum bits ``qubits'' (on the order of 1000)
would be able to solve certain problems
that are intractable on classical computers.
Building such a device is a formidable task, and several
radically differing designs have been proposed \cite{nielsen00}.
One promising approach is to encode quantum information
using the spin of single electrons confined in semiconductor quantum dots
\cite{loss98a,burkard99a,hu00a,friesen03a}.
Universal quantum computation \cite{divincenzo95b,barenco95a}
in this approach uses a tunable kinetic exchange interaction
between the dots (resulting in a Heisenberg interaction) 
and one-qubit rotations, which can be obtained by applying local magnetic
fields in at least two directions.

The one-qubit rotations
are much more difficult to control experimentally than
the kinetic exchange interaction.
This spurred a number of proposals of quantum computation schemes
using the exchange interaction alone
\cite{divincenzo00a,kempe01b,levy02a,lidar02a,bacon00a,lidar00a}.
To use this single interaction,
the quantum information must be encoded in multiple (two or more)
spins.

Decoherence due to interactions with the environment
pose a much larger problem for qubits than for classical
bits, and there has been a tremendous effort on developing ways
of protecting quantum information from
decoherence \cite{bacon00a,lidar00a,gottesman96a,knill97a,calderbank97a,duan97a,kane98a,burkard99b,steane99a,viola00a,ioffe02a,meier03a,wu02a}.
To shield quantum information from the environment,
Zanardi and Rasetti \cite{zanardi97a} first proposed encoding
quantum information in in the ``noiseless'' singlet
subspace of 4 (or more) 2-level systems.
This subspace, often called 
a Decoherence Free Subspace (DFS),
is immune to collective decoherence,
that is, environment-induced dephasing that acts equally
on each constituent element of the composite qubit 
\cite{bacon00a,lidar00a,kielpinski01a,lidar98a,lidar99a,kempe01a,lidar03a}.

Bacon, Brown, and Whaley (BBW) proposed
creating a supercoherent qubit from the 4-spin DFS
by generating an energy gap from the singlet subspace
to the rest of the Hilbert space
\cite{bacon01a}.
Decoherence must overcome this energy gap and is suppressed
exponentially for temperatures much smaller than this 
gap \cite{barnes00a}.
To create the gap, BBW use equal antiferromagnetic
interactions between all pairs of four $s=1/2$ spins.
The Hamiltonian is simply
${\rm \bf H} = J\sum_{i<j} {\rm \bf S}_i \cdot {\rm \bf S}_j $,
and 
the eigenvalues 
$E(S) = (J/2) ( S (S+1)-3) $
depend only on the total spin $S$.
The quantum information is encoded in the doubly degenerate singlet ground 
state.

Building four quantum dots with equal tunneling rates between
each pair is challenging.  
Tunneling rates decay exponentially with the separation between dots.
Arranging the dots 
on the vertices of a three-dimensional tetrahedron
would work, but it is far preferable to build two-dimensional structures.
A square arrangement will have
much weaker tunneling across the diagonal
of the square than along the edges of the square.

\begin{figure}[tbh]
\begin{center}
\includegraphics[width=1.2in]{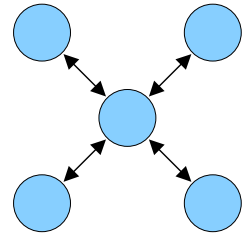}
\end{center}
\vspace{.1in}
\caption{The proposed configuration of 5 quantum dots
in a two-dimensional plane.
The four tunnelings between the outer dots and the middle dot
are always on.
This system with 4 (or 6) electrons has a doubly degenerate
singlet ground state.
}
\label{fig:five}
\end{figure} 

\begin{figure}[t]
\begin{center}
\includegraphics[width=3.375in]{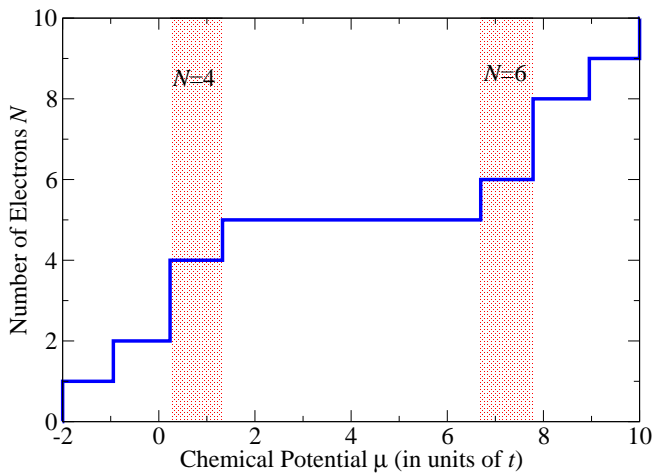}
\end{center}
\vspace{.1in}
\caption{
Total occupancy $N=\sum_i n_i$ of the 5-dot system as a function of overall
chemical potential $\mu$.
Here $t_{ij} = 1$, $U_i=8$, and $\mu_i=\mu$.
The ground state is a doubly degenerate singlet when the system
is filled with 4 or 6 electrons.
}
\label{fig:grand}
\end{figure} 

We propose adding an extra dot (and not an extra electron)
in the middle of a 
square arrangement of four dots, as shown in Fig.~(\ref{fig:five}).
We separate the outer four dots so the direct tunneling between
them is negligible.
Since the ground state wave function
of the middle dot will have $s$-wave character,
the effective interaction between each pair of the outer four dots
will be equal if the four tunnelings between the outer and middle 
dots are made equal (e.g.\ by tuning gates located above or below
each tunneling region).

This design is inspired by the superexchange process,
which uses empty (or filled) auxiliary quantum dots
to mediate interactions between 
dots that are too widely separated to interact directly
\cite{recher00a,manousakis02a,auerbach94}.
Electrons (or holes) can reach distant quantum dots
by hopping through the auxiliary dots.

We need to verify that four electrons can be placed
in the five-dot system.
We model the system with a Hubbard Hamiltonian using one
orbital per quantum dot:
\begin{eqnarray}
 H & = & \sum_{i,j,\sigma } t_{ij}
  c_{ i \sigma}^\dagger c_{ j \sigma}
+  \sum_i \left( U_i n_{i\uparrow} n_{i\downarrow} - \mu_i n_i \right) 
\end{eqnarray}
where $t_{ij}$ is the hopping amplitude between dots $i$ and $j$,
$U_i$ is the Coulomb repulsion between two electrons on dot $i$,
$ n_i = n_{i\uparrow} + n_{i\downarrow}$ is the total number
of electrons on dot $i$,
and $\mu_i$ is the onsite potential of dot $i$ \cite{burkard99a}.
The calculations used $t_{ij} = -1$, $U_i=8$, and an equal overall
potential on each dot of $\mu_i=\mu$.
The total occupancy of the ground state of the system calculated
by exact diagonalization in the grand canonical ensemble is shown 
in Fig.~(\ref{fig:grand}).
The largest region of stability contains five electrons, which
is to be expected in the large $U$ limit, but
there are significant ranges of the chemical potential $\mu$
for which the ground state has four and six electrons.
In these regions, the ground state is a doubly degenerate singlet,
and quantum computation in this 
supercoherent subspace
is possible.

\begin{figure}[tbh]
\begin{center}
\includegraphics[width=3.375in]{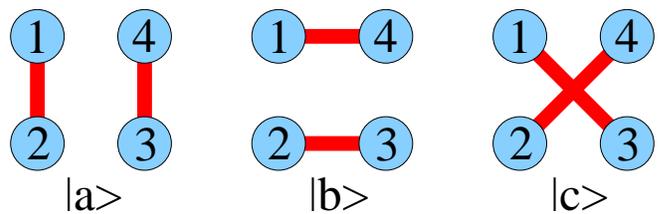}
\end{center}
\vspace{.1in}
\caption{Ground states of the 5-dot composite qubit.
In each eigenstate, the middle dot is omitted.
The lines connecting dots represent singlet valence bonds when both
dots are occupied.
These three states are not orthogonal: two orthogonal states may be formed,
for example, as 
$|0\rangle = |{\rm a}\rangle$ and
$|1\rangle = (|{\rm b}\rangle + |{\rm c}\rangle)/\sqrt{3}$.
}
\label{fig:eigenstates4}
\end{figure} 

The Hamiltonian used to generate Fig.~(\ref{fig:grand}) has equal onsite
potentials at every dot, and thus has 
electron-hole symmetry.
Raising the potential of just the middle dot breaks this symmetry
and increases the range of
chemical potentials yielding the $N=4$ ground state
while lowering this potential increases the range of the $N=6$
ground state.
Quantum computation is possible using either the $N=4$ or $N=6$
ground states, and the gate operations for the two cases are identical.

A simple way of describing
eigenstates of the 5-dot composite qubit is shown in 
Fig.~(\ref{fig:eigenstates4}) using
valence-bond representation \cite{auerbach94}.
The ground states contain two separate singlet bonds.
There are three ways to construct these bonds:
Dot number 1 is bonded to any of the other three outer dots,
and then the other bond is formed between the two outer dots
not bonded to dot 1.
Valence-bond states are not orthogonal in general,
and two orthogonal states may be formed from the three states
in Fig.~(\ref{fig:eigenstates4}).
There are several ways to form two orthogonal states, and
we will use
$|0\rangle = |{\rm a}\rangle$ and
$|1\rangle = (|{\rm b}\rangle + |{\rm c}\rangle)/\sqrt{3}$.

The information encoded in the degenerate total-spin singlet subspace
of the 5-dot composite qubit is immune to collective decoherence,
that is, decoherence affecting all spins 
equally \cite{bacon00a,lidar00a,zanardi97a,kielpinski01a,lidar98a,lidar99a,kempe01a,lidar03a,bacon01a}.
It is also immune to local decoherence affecting only a single dot.
To see this, consider a magnetic field or an extra electron coupling only to
dot 1 in Fig.~(\ref{fig:eigenstates4}).
The singlet bond connecting to spin 1 will be mixed with a triplet bond.
However, this occurs equally to all three eigenstates in
Fig.~(\ref{fig:eigenstates4}),
so the degeneracy between these states is not broken.
More complicated mechanisms affecting multiple spins can cause
decoherence, but these mechanisms must overcome the energy gap to the
first excited state.
 
We now demonstrate the physically intuitive
gate operations on the 5-dot qubit that 
allow universal quantum computation.
Notice that
$|0\rangle = |{\rm a}\rangle$ in Fig.~(\ref{fig:eigenstates4})
is odd under an
exchange of sites 1 and 2, denoted by $1 \leftrightarrow 2$,
and this state is also odd under $3 \leftrightarrow 4$.
The other ground state
$|1\rangle = (|{\rm b}\rangle + |{\rm c}\rangle)/\sqrt{2}$
is even under both operations.

\begin{figure}[tbh]
\begin{center}
\includegraphics[width=1.2in]{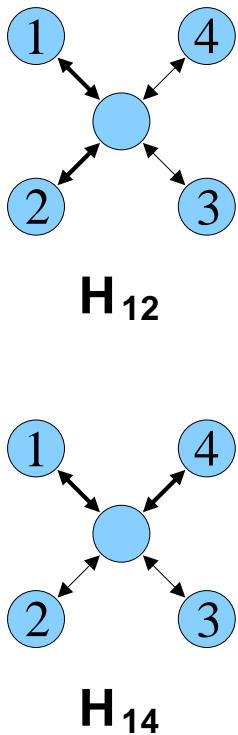}
\end{center}
\vspace{.1in}
\caption{
Single-qubit rotations are performed by varying
at least two of the tunneling parameters, shown by thicker lines.
In H$_{12}$, for example, the tunnelings between the 
the central dot and dots 1 and 2 are increased
relative to the tunnelings between the central dot
and dots 3 and 4.
H$_{12}$ splits but does not mix states $|0\rangle$ and
$|1\rangle$, and thus functions as a field in the $\hat{z}$
direction in pseudospin space.
H$_{14}$ splits and mixes the states as described in the text.
Combinations of H$_{12}$ and H$_{14}$ allow arbitrary
$SU$(2) rotations of the single composite qubit.
}
\label{fig:gate1}
\end{figure} 

Increasing the tunneling between the middle dot and dots 1 and 2,
denoted by $H_{12}$ in Fig.~(\ref{fig:gate1}),
respects the $1 \leftrightarrow 2$ and  $3 \leftrightarrow 4$ symmetries
and does not mix the ground states, but $H_{12}$ does break the
degeneracy of the ground state.
Thus it acts in the pseudospin space of $|0\rangle$ and $|1\rangle$ as
a magnetic field in the $\hat{z}$ direction:
\begin{equation}
H_{12} \propto \left(
\begin{array}{rr}
1 & 0 \\
0 & -1 
\end{array}
\right) .
\label{h12}
\end{equation}
The tunnelings from the middle dot to dots 3 and 4
could have been increased to implement the same gate,
so effectively $H_{34} = H_{12}$.
Adiabatic variation of the tunneling rates is required
to avoid mixing with excited states.
Adiabaticity is also required in the conventional
single-quantum-dot qubit implementation to avoid mixing
with states containing dots occupied by two 
electrons \cite{loss98a,hu00a,schliemann01a}.

To perform arbitrary $SU$(2) rotations of the single composite qubit, 
we need to be able to perform rotations about two directions
in pseudospin space.
Therefore a gate in addition to $H_{12}$ is required.
Varying three tunnelings in the 5-dot composite qubit can produce
a rotation about the $\hat{x}$ axis in pseudospin space,
but a simpler gate can be formed by increasing the
tunneling between the middle dot and dots 1 and 4,
denoted by $H_{14}$ in Fig.~(\ref{fig:gate1}).
This operation breaks the $1 \leftrightarrow 2$ and  
$3 \leftrightarrow 4$ symmetries,
and can be shown to be
\begin{equation}
H_{14} \propto \left(
\begin{array}{rr}
-1 & \sqrt{3} \\
\sqrt{3} & 1
\end{array}
\right) ,
\label{h14}
\end{equation}
which represents a rotation at an angle of 120 degrees from
the $\hat{z}$ axis \cite{divincenzo00a,bacon00a}.
Again, increasing the tunnelings to dots 2 and 3 would implement the
same gate, so $H_{23} = H_{14}$.
With $H_{12}$ and  $H_{14}$, any $SU$(2) rotation can be performed
on the composite qubit.
The one-qubit operations are similar to operations described
for the 4-spin DFS, but in that case the interaction takes
place directly between the spins and not through an auxiliary
fifth dot \cite{bacon00a,lidar00a,kempe01a}.

\begin{figure}[tbh]
\begin{center}
\includegraphics[width=3.1in]{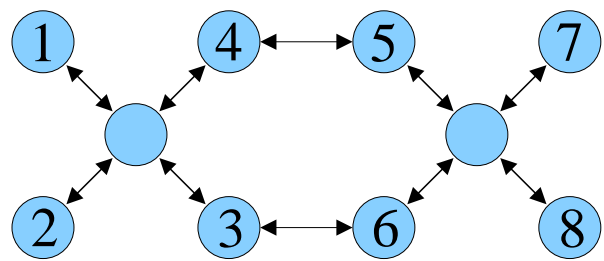}
\end{center}
\vspace{.1in}
\caption{
A two-qubit operation can be performed by turning on tunneling between
two pairs of dots in neighboring composite qubits.
Combined with the single-qubit rotations in Fig.\ 4, this
operation allows any arbitrary unitary transformation to
be performed.
}
\label{fig:gate2}
\end{figure}

To form a two-qubit gate, two tunnelings must be
turned on between adjacent qubits---A single tunneling interaction
performs no operation due to the
immunity of the composite qubit to local decoherence.
An example of a two-composite-qubit gate is shown in Fig.~(\ref{fig:gate2})
in which tunneling between pairs of outer dots on adjacent qubits
has been turned on.
This operation preserves the symmetries $1 \leftrightarrow 2$
and $7 \leftrightarrow 8$.
Thus its action in the basis of 
$\{|00\rangle,|01\rangle,|10\rangle,|11\rangle\}$  has
the general form
\begin{equation}
H_{\rm 2 qubit} = \left(
\begin{array}{cccc}
A & 0 & 0 & 0 \\
0 & B & 0 & 0 \\
0 & 0 & B & 0 \\
0 & 0 & 0 & C 
\end{array}
\right) 
\label{gate2}
\end{equation}
which has been verified by exact diagonalization of the 10-dot 8-electron
Hubbard model for the two-dot system.
During both the single- and two-composite qubit gate operations,
the system stays in the total singlet subspace and remains
immune to collective decoherence \cite{bacon00a}.

Combining $H_{\rm 2 qubit}$
with single qubit rotations on the individual dots
allows the controlled-phase gate $\bar{\bf C}_P$ to be implemented with
a single pulse in which six tunneling rates are varied:
\begin{eqnarray}
\bar{\bf C}_P
& = & \exp{ \left( \frac{i}{\hbar} \int 
\left (H_{12}(t) + H_{\rm 2 qubit}(t) + H_{78}(t) \right) \right)
{\rm d}t}
\nonumber
\\
& = & \left( \begin{array}{rrrr}
1 & 0 & 0 & 0 \\
0 & ~~1 & 0 & 0 \\
0 & 0 & ~~1 & 0 \\
0 & 0 & 0 & -1
\end{array}
\right)
\end{eqnarray}
where $H_{12}$, $H_{\rm 2 qubit}$, and $H_{78}$ all 
commute \cite{nielsen00,bacon00a,kempe01a}.

Finally, it is interesting to note that the 
5-dot configuration is actually more stable than a 4-dot setup to variations 
in the hopping parameters.
Varying a single tunneling rate in Fig.~(\ref{fig:five})
does not break the degeneracy of the ground state.
Such a change modifies the effective interactions of a single
outer dot with each of the other outer dots, and simply 
shifts the ground-state energy preserving the degeneracy.
In contrast, varying one of the six tunneling rates in 
the 4-dot setup
splits the ground states.

In summary, a five-dot composite qubit design was presented that operates in a
decoherence-free subspace.
Universal quantum computation is easily implemented by
varying tunneling rates in a simple, physically intuitive manner---Generation
of local magnetic fields is not required to perform the gate operations.
Each qubit is encoded in the degenerate singlet ground-state
of four (or six) electrons
in a system of five quantum dots arranged in a two-dimensional pattern.
This 
supercoherent 
subspace
is immune to both collective and local decoherence,
and resists other forms of decoherence, which must raise the energy.

We thank J. Levy and D. A. Lidar for stimulating discussions.
Support 
from
DARPA QuIST (MIPR 02 N699-00) 
and DOD HPCMO CHSSI
is gratefully acknowledged.

\end{document}